\documentclass{article}

\begin{document}

\title{The Hole Argument for Covariant Theories}
\author{Iftime, M.\thanks{School of Arts and Sciences, MCPHS, Boston, MA, USA} , \, Stachel, J. \thanks{ Department of Physics and Center for Philosophy and History of Science, Boston, MA, USA}}
\date{}
\maketitle

\begin{abstract}

The hole argument was developed by Einstein in 1913 while he was searching for a relativistic theory of gravitation.
Einstein used the language of coordinate systems and coordinate invariance, rather than  the language of manifolds and diffeomorphism invariance. He formulated the hole argument against covariant field equations and later found a way to avoid it using coordinate language. 

In this paper we shall use the invariant language of categories, manifolds and natural objects to give a coordinate-free description of the hole argument and a way of avoiding it. Finally we shall point out some important implications of further extensions of the hole argument to sets and relations for the problem of quantum gravity. 

\end{abstract}


\pagebreak

\tableofcontents

\pagebreak

\section{The Hole Argument in General Relativity}

\subsection{The Original Hole Argument}

In Einstein's most detailed account of the hole argument \cite{Einstein1914}, $G(x)$ represents a metric tensor field that satisfies the field equations in the $x$-coordinate system and $G'(x')$ represents the same gravitational field in the $x'$-coordinate system. Einstein realized that, if the field equations for the metric tensor are covariant, then $G'(x)$ must also represent a  solution to these equations in the $x$-coordinate system. He asked the following question: 
Do the metrics $G(x)$ and $G'(x)$ represent the same or distinct gravitational fields ? 
If they represent distinct gravitational fields, as Einstein originally assumed, then the hole argument showed that no specification of the metric field outside of and on the boundary of a hole ( i.e. an open set of space-time) could uniquely determine the field inside the hole. 

Einstein posed the hole argument as a boundary-value problem. Hilbert 
realised that it was more appropriate to formulate it as an initial value problem.\footnote{For modern discussions of the implications of the diffeomorphism invariance for the Cauchy problem in general relativity, see \cite{HE}, \cite{Geroch}.}
If $G(x)$ and $G'(x)$ represent distinct solutions, then the specification of the $G$-field and any finite number of its time derivatives on an initial hypersurface $t=0$ could determine the field uniquely of the initial hypersurface. These two formulations are refered as the boundary-value and initial-value formulations of the original hole argument.  \footnote{For more details and a historical review of the original hole argument see \cite{Stachel1989}, \cite{Stachel1993}.}

We shall formulate the hole argument in manifold language. Our approach will be global, in the sense that the whole base (space-time) manifold $M$ and (global) diffeomorphisms of $M$ come into play, rather than working in a coordinate patch and local coordinate transformations.

A general relativistic space-time is a $4$-dimensional manifold $M$, together with 
a Lorentz metric field $g$ on $M$.\footnote{Time-orientability is an extra condition needed to formulate the global Cauchy problem and global causal structure on space-time \cite{HE},\cite{Klainerman}.} 
This metric field represents not only the chrono-geometrical structure of space-time, but also the potentials for the inertio-gravitational field.\footnote{See e.g., \cite{Stachel1994}, 
\cite{Stachel2000}.} 

Einstein's field equations are {\em covariant}. This means that, if $g$ is a solution of Einstein's equations, then all pull-backs metrics $\phi^{*}(g)$ obtained from $g$ by the mapping induced by diffeomorphisms $\phi$ on $M$ also satisfy Einstein's equations. 

The question at the heart of the hole argument is the following. 
 
Do all metrics $\phi^{*}(g)$ describe the same gravitational field?  
Einstein's ultimate answer was ''yes''. This assertion is what we call {\em general covariance} of Einstein's equations. \footnote{Some authors define general covariance as what we call covariance, not making a clear distinction between the two. See \cite{Stachel1993} for this distinction.}

Let us suppose a space-time manifold $(M,g)$ contains a {\em hole}, i.e., an open space-time region $H$ on which the metric field $g$ is the only one present, so that inside $H$, the metric $g$ obeys the (homogeneous) Einstein empty space-time equations:

\begin{equation}
Ein(g)=Ric(g) -\frac{1}{2}gR(g)=0
\end{equation}

Given a solution $g$ everywhere outside of and on the boundary of the hole $H$, including all the normal derivatives of the metric up to any finite order on that boundary, the metric field $g$ inside $H$ is still {\em not determined uniquely} no matter how small the hole. 
The proof: Given any solution inside the hole $H$, an 
unlimited number of other solutions can be generated from it by those diffeomorphisms that 
reduce to the identity on $M-H$ (and any number of derivatives of which also reduce to the 
identity on the boundary), but differ from the identity inside $H$.
This brings us to the following contradiction (the hole argument): No well-posed initial-value and/or boundary-value problem can be posed for Einstein's covariant equations\footnote{ The proof works also for any set of covariant equations} So, such field equations would seem not to be of much use, which is why Einstein (and then Hilbert) initially rejected them. 
\cite{Stachel1999}

Einstein found a way to evade the hole argument using coordinate language. Translated in the language of manifolds, one must assume that, at least inside the hole $H$, the points of the manifold are {\em not} individuated independently of the metric field. 
This means that space-time points have {\em no inherent} chrono-geometrical or inertio-gravitational properties or relations that do not depend on the presence of the metric tensor field. This implies that when we drag-along the metric, we actually drag-along the physically individuating properties and relations of the points. Thus, the pull-back metric does not differ {\em physically} from the original one. It follows that the entire equivalence class of diffeomorphically-related solutions to Einstein's empty space-time field equations corresponds to {\em one} inertio-gravitational field.

Put in other words, while the points of the manifold have an inherent {\em quiddity} as elements of space-time, they lack {\em haecceity} \footnote{For a discussion of {\em quiddity} and {\em haecceity}, see \cite{Stachel2004}, p. 204 and \cite{Stachel2005}} as individualized points of that space-time (''events'') unless and until a particular metric field is specified.
\footnote{In the generic case (i.e. no symmetries present), 
the four non-vanishing invariants of the Riemann tensor in empty space-times can be 
used to individuate the points of space-time. See \cite{Stachel1993}, pp. 155-156} 

Mathematically, we have $\mathcal{M}(M)$ the collection of all Lorentz metrics on a space-time manifold $M$ and $Diff(M)$ the group of diffeomorphisms of $M$. 

$Diff(M)$ acts as a transformation group on $\mathcal{M}(M)$ by pulling-back metrics on $\mathcal{M}(M)$: for all $\phi\in Diff(M)$ and $g\in \mathcal{M}(M)$ the action map is defined by $(\phi, g)\longmapsto \phi^{*}(g) $. 

For a fixed metric $g$, $\mathcal{O}_{g}=\{\phi^{*}(g)|\quad \phi\in Diff(M)\}$ is the orbit through $g$. Two metrics $g_1$ and $g_2$ are on the same orbit, if and only if $\phi^{*}(g_1)=g_2$ for some  $\phi\in Diff(M)$, i.e. $g_1$ and $g_2$ are isometric.

If $g_1$ satisfies Einstein's equations  then $g_2$ does. Two space-time solutions of Einstein's equations are considered to be {\em physically equivalent} if one is isometric to the other. 

The action of $Diff(M)$ on $\mathcal{M}(M)$ partition $\mathcal{M}(M)$ into (disjoint)
isometry classes of Lorentz metrics.

The space $\mathcal{Q}(M)$ of all isometry classes of space-time metrics on $M$ is the quotient $\mathcal{M}(M)/Diff(M)$. A physical space-time is therefore a point in 
$\mathcal{Q}(M)$.  The space $\mathcal{Q}(M)$ does not have a manifold structure, 
but in some cases it can be shown that $\mathcal{Q}(M)$ forms a stratified manifold, 
with each strata consisting of space-times with conjugate isometry groups. \cite{Fisher}, \cite{Marsden} One of the reasons why $\mathcal{Q}(M)$ does not have a manifold structure is because the isometry (symmetry) group  $\mathcal{H}_{g}=\{\phi \in Diff(M) | \phi^{*}(g)=g\}$ is not trivial, if the metric $g$ has symmetries.

The projection map $\Pi: \mathcal{M}(M)\longrightarrow \mathcal{Q}(M)$ identifies all isometric Lorentz metrics to a single diffeomorphically-equivalence class. The image of $\Pi(g)=[g]$ represents the (unique) physical gravitational field defined by $g$.

\subsection{The Hole Argument for Inhomogeneous Einstein's Field Equations}

Let $(M, g)$ be a space-time manifold. There may also be a collection $\psi $ of other tensor fields on $M$, representing non-gravitational fields and/or matter and acting as sources of the metrical field in the inhomogeneous Einstein's equations:

\begin{equation}
Ein(g)=Ric(g) -\frac{1}{2}gR(g)=T(g,\psi)
\end{equation}

In most cases the metric field appears on both left and right sides of Einstein equations, so the space-time structure and the source fields in space-time 
constitute a dynamical system, the equations of which can only be solved together.\footnote{ For some exceptions, see \cite{Stachel1969}}

It is clear that a generalized version of the hole argument can also be applied to regions $H$ of a space-time $M$, in which the inhomogeneous Einstein equations hold, together with the set of dynamical equations obeyed by the non-gravitational matter and fields( e.g. coupled Einstein-Maxwell or  Einstein-Yang Mills equations), provided this set of coupled gravitational and non-gravitational field equations has the {\em covariance} property. (The generalization is straightforward and we leave the details to the reader.)

In order to avoid this version of the hole argument, we must assume that space-time points of this region $H$ are not individuated unless and until both the gravitational and non-gravitational fields are specified in $H$. \footnote{This does not imply that {\em all} of these fields are 
necessary for such individuation. The values of four independent invariants of the metric fields will suffice to individuate the space-time points in the generic case.}  

Thus, an entire class of diffeomorphically-related solutions to the coupled equations 
will correspond to one set of physical gravitational and non-gravitational fields.

\section{The Hole Argument for Covariant Theories}

We shall formulate the hole argument to theories that are covariant and give a way of avoiding it.

\subsection{Theory, Models, Covariance and General Covariance}

At the level of physics \footnote{or indeed any other natural science that has reached the level of abstraction, at which mathematical structures may be usefully correlated with the concepts of this science}, we have the following correlations between physical concepts and mathematical structures:

{\em A type of theory} is correlated with the concept of a type of {\em natural bundle} or 
{\em gauge-natural bundle}.

A  natural bundle is defined as a covariant functor $\mathtt{F}$  of the category whose objects are $n$-dimensional manifolds and whose morphisms are local 
diffeomorphisms into the category whose objects are fibred manifolds and 
whose morphisms are fibre-preserving morphisms. 
We can think of $\mathtt{F}$ as a rule such that it takes an $n$-dimensional manifold $M$ into a fibered manifold $(\mathtt{F}M=E\stackrel{\pi}{\longrightarrow}M)$ and a local diffeomorphism  $\phi :M\to N$ between two $n$-dim manifolds $M$ and $N$ to a fiber-preserving morphism $\mathtt{F}\phi$ over $\phi$.

A gauge-natural bundle can be defined similarly as a 
covariant functor $\mathtt{F}$  from the category of principal bundles and 
principal fibre-preserving morphisms into the category of fibred manifolds and fibre-preserving morphisms.

Notationally, we will make no distinction between the functor $\mathtt{F}$, the fibered manifold 
$(E\stackrel{\pi}{\longrightarrow }M)$ or the total space $E$, but one should keep in mind that $F$ is a functor, i.e. the object $(E\stackrel{\pi}{\longrightarrow }M)$ together with a class of automorphisms. 

{\em A mathematical model} of a type of theory is correlated with a {\em cross-section} of the natural bundle $(E\stackrel{\pi}{\longrightarrow }M)$, or a class of gauge-equivalent cross sections of the gauge-natural bundle. 

A {\em particular theory of given type} is correlated with a rule for selecting a class of cross-sections of the (gauge-)natural bundle, that is a class of mathematical models of that type of theory. 

For example, Maxwell's electromagnetic theory is a
rule for selecting the class of cross-sections of the gauge-natural bundle of
one-form fields that obey the linear, gauge-invariant field equations derived from the Maxwell
lagrangian. On the other hand, Born-Infeld electrodynamic theory is a rule for selecting a (different) class of cross-sections of the gauge natural bundle of one-form fields that obey the
non-linear, but gauge-invariant field equations derived from the Born-Infeld lagrangian 
\cite{BornInfeld}. 

As these examples illustrate, for a theory based upon field equations derived from the variation of a lagrangian, the selected cross-sections must satisfy some set of differential equations for the type of geometric object. 

So, in order to formulate the rule of selecting cross-sections one need to use jet extensions of the configuration space \cite{Sardanashvily}. But for our purposes we do not need go to any further details about the rule for selecting cross-sections(see \cite{Hermann}, Chapter I, pp. 1-15 and  \cite{KMS1993}, Chapter IV,  pp. 124-125).

General relativity can be formulated in terms of natural bundles.
\footnote{There two major ways to formulate general relativity; one uses only the metric and the second order jet extension (\cite{Sachs}, \cite{HE}) and the other, the metric and the connection and the first order jet extension \cite{Geroch}.}
A mathematical model consists of a cross-section $\sigma$ of the associated bundlle $\Sigma _{PR}$ of the linear frames $LM$ with standard fiber $GL(4)/O(1,3)$. \cite{Sardanashvily}

In most cases the metric must satisfy some additional criteria, e.g. timelike and/or null geodesic completeness. In general, the cross-sections $\sigma$ can be viewed as a maximal extension of a local cross-section subject to some additional criteria. 

The properties of local cross-sections of a fiber bundle may be expressed in the language of sheaf theory. In terms of sheaves, giving a mathematical model means giving a sheaf of local cross-sections of the fibered manifold $(E\stackrel{\pi}\to M)$.  Alternatively, we can replace the sheaf of cross-sections by its corresponding {\'{e}}tale bundle $(\hat{E}\stackrel{\hat{\pi}}\to M)$ over $M$. \cite{MacLane1992}

The following question arises: Do all mathematically distinct models of a physical theory necessarily correspond to physically distinct models of the theory? 
As we shall see later, the hole argument concerns this question. The answer to this question is the basis for the distinction between covariant and generally covariant theories.

\subsection{Background Independent vs Background Dependent Theories}

A {\em background independent theory} is a physical theory defined on a base manifold $M$ endowed with no extra structure, like geometry or fixed coordinates.
If a theory does include any such geometric structures, it is called a {\em background-dependent}.

Examples of background dependent theories are all special-relativistic field theories: the Minkowski metric and the (unique) symmetric inertial connection compatible with it are background structures of these theories. More generally, a field theory may be specified in a space-time with any given background Lorentz metric and the corresponding (fixed) inertio-gravitational connection. One may say that, in any such background-dependent theory, there is a kinematics that is {\em prior} to and independent of the dynamics of the theory.

Examples of background independent theories are all general-relativistic gravitational theories, obeying either the Einstein empty space-time equations or a set of generally covariant coupled equations for a theory of matter fields coupled to Einstein's gravity.\footnote{In the case of a background independent classical theory, it seems natural to require background independence in the corresponding quantum theory. See, for example
\cite{Lewandowski} for a discussion on background independence quantization of a scalar field} 
Both the Lorentz and  the compatible inertio-gravitational connection are among the dynamical variables, for which a solution to the Einstein's equations must be specified. One may say that, in a background independent theory, there is no kinematics prior to and independent of the dynamics of the theory. It is to such background independent theories that the hole argument applies.

\subsection{The Hole Argument for Geometric Objects}

Let us consider a particular theory $\mathcal T$ defined on a manifold $M$ i.e. a rule for selecting a set of models of a type of theory $(\mathtt{F}M=E\stackrel{\pi}\to M)$. In other words we pick up a sheaf $\Gamma(\pi) $ of local cross-sections of ${\pi}$.

If $\phi: M\to M$ is an arbitrary diffeomorphism of the base manifold $M$, from the definition of a natural bundle follows that there is a uniquely defined fibered manifold automorphism 
$\mathtt{F}\phi=\Phi: E\to E$ that projects over $\phi$ i.e.,  $\Phi$ is diffeomorphism of $E$ and $\pi\circ \Phi=\phi\circ \pi$. Such a fibered manifold automorphism  $(\Phi,\phi)$  can be used to transport local cross-sections of $\pi$ to local cross-sections of $\pi$:  

If $\sigma$ is a local cross-section of $\pi$ defined on a small open set $U_p$ around $p\in M$, then $\phi^{*}\sigma$ defined by
$\phi^{*}\sigma =\Phi\circ\sigma\circ\phi^{-1}$ is a new local cross-section of $\pi$
defined on the open set $U_{q}=\phi(U_{p})$ around $q=\phi(p)\in M$, and 
called the {\em carried-along cross-section} of $\sigma$ by the fibered manifold automorphism 
$(\Phi, \phi)$.

In functorial language, a diffeomorphism $\phi$ will induce functors in both directions on the associated categories of sheaves of cross-sections on $M$.  To each sheaf of cross-sections $\Gamma(\pi) $, $\phi^{*}\Gamma(\pi)$ is {\em the direct image} of the sheaf of cross-sections $\Gamma(\pi)$ under the base diffeomorphism $\phi: M\to M$:

\begin{equation}
\phi^{*}\Gamma(\pi)(U_{q})=\Gamma(\pi)(U_{p}),\qquad U_{p}=\phi^{-1}_{M}(U_{q})
\end{equation}

If $\sigma\in \Gamma(\pi)(U_{p})$, then the carried-along cross-section 
$\phi^{*}\sigma\in \phi^{*}\Gamma(\pi)(U_{q})$. 

In the case  when $E:=TM$ is the tangent bundle over $M$, if $\phi: M\to M$ is an arbitrary diffeomorphism on $M$, then $(\phi_{*},\phi_M)$ is a bundle isomorphism of $TM$. So, if
$v$ is a vector field on $M$, then $\phi_{*}v $ is the carried along tangent vector field. In the case of the metric field, $\phi^{*}g$ is the pulled-back metric of $g$ by the adjoint map of $\phi_{*}$, where $(\phi^{*},\phi)$ is a bundle isomorphism of the cotangent bundle $T^{*}M$.

A theory $\mathcal T$ is {\em covariant} if all the carried-along cross-sections $\phi^{*}\sigma$ of any model $\sigma$ in the theory are also models of the theory. 
Two models related by such a local diffeomorphism are called {\em diffeomorphically-equivalent}. This relation is clearly an equivalence relation and it divides all models of $\mathcal T$ into diffeomorphically-equivalent classes of cross-sections on $M$.  

A covariant theory $\mathcal T$ is called {\em generally covariant} if 
all the carried-along cross-sections $\phi^{*}\sigma$ of any model 
$\sigma$ in the theory represents the {\em same} physical model. 

It is the presence of non-dynamical individuating fields  that prevents a covariant theory from being generally covariant, enabling one to distinguish between two models in the same equivalence class.\footnote{ E.g., the color space manifold \cite{Stachel1989}.}

The original hole argument generalized to the case of geometric objects translates in the possibility that the two models have physically identical properties.

Let $\mathcal{T}$ be a covariant  theory on a base manifold $M$. Let $H$ be a ''hole'',  which is an open set in $M$.  Giving a model $\sigma$ 
everywhere outside of and on the boundary of the hole $H$, we cannot uniquely determine $\sigma$ inside $H$ no matter how small the hole in $M$. From the covariance property, the hole argument hinges on the answer to the following question: 

Is it possible to pick out a {\em unique model} within an equivalence class by specifying $\sigma$ everywhere on $M$ except on some open submanifold $H\subseteq M$ ? \\
The proof (that the answer to this question is negative) is analogous with that in the original argument. Given a model $\sigma$ on $M$, an unlimited number of other solutions $\phi^{*}\sigma$  that are identical with $\sigma$  on $M-H$ and on the boundary of $H$, but differ inside $H$, can be generated from $\sigma$ by those diffeomorphisms of $M$ that reduce to the identity on $M-H$ on the boundary of $H$,  but differ from the identity inside $H$. Then $\sigma$ and $\phi^{*}\sigma$
will be two different models that agree on $M - H$ but differ on $H$ \footnote{Of course, they belong to the same equivalence class, but they are not automatically physically equivalent}.  
In this case, no conditions imposed on $\sigma$ on $M -H$ can serve 
to fix a unique model on $H$, no matter how small the hole is. The only way to specify such a model uniquely, is to specify $\sigma$ everywhere on $M$. 
This is the hole argument for covariant theories, or rather against them.

If we postulate that each diffeomorphically-equivalent class of cross-sections corresponds to 
{\em one} physical solution, then this requirement depends on the assumption that the base points of $M$ are individuated exclusively by the model (i.e., the class of cross-sections).
In other words, the hole argument validity depends crucially on the assumption 
that the distinction between the points of the manifold (i.e., their haecceity),
is {\em independent} of the specification of a particular model of the theory $\mathcal T$. 

If the individuation of the points of $M$ depends entirely on the model, then
we have no grounds for asserting that $\sigma$ and $\phi^{*}\sigma$ represent different physical models. Given the lack of any model-independent distinction between the points of 
the manifold, no distinction can be made between the models $\sigma$ and $\phi^{*}\sigma$, so they represent the same physical model. Conversely, if $\sigma$ and  
$\phi^{*}\sigma$ represents the same physical model for all base diffeomorphisms $\phi$, then the hole argument clearly fails. It follows that (as far as concerns the covariant 
theory $\mathcal T$ under consideration) the points of $M$ must be entirely unindividuated before a model $\sigma$ is introduced. All relevant distinctions between these points must be consequences of the choice of the model $\sigma$, i.e. such a theory is generally covariant. 

Any background independent theory is general covariant. For background independent theories (e.g., general relativity), the gauge freedom is the full diffeomorphism group $Diff(M)$ of the base manifold $M$: under any arbitrary base difeomorphism $\phi$, the two mathematical models$(M,\sigma)$ and $(\phi(M)= M, \phi^{*}\sigma)$ are physically 
indistinguishable. In coordinate language, the coordinates (or physical components) of $\phi^{*}\sigma$ in the carried along frame are the same as the coordinates of in the original frame. Informally this means that {\em if everything is carried along, nothing is changed}. \footnote{The two models 
$(M, \sigma)$ and $(M, \phi^{*}\sigma)$ are {\em elementarily equivalent} models in the sense of model theory i.e., they share the same model-theoretic properties \cite{Hodges}. The truth values or the probability of the corresponding assertions in each model will always be the same. That is for every assertion about the model $(M, \sigma)$, there is a one-to-one corresponding assertion about models $(M,\phi^{*}\sigma)$.}

The hole argument can only apply to background-independent theories. But in between (full) background independence and (full) background dependence (i.e., maximal symmetry group) there are fields with non-maximal symmetry groups and non-trivial orbits. These case have to be treated by examining the action of the full diffeomorphism group on the quotient space.
\cite{MaCCallum}

If the theory is background dependent then the hole argument fails automatically. 
A background space-time structure is sufficiently ''rigid'' to prevent the hole argument formulation.  Specification of a solution outside of and on the boundary of a hole $H$ completely determines, indeed over-determines a solution inside $H$. The class of physical fields on $M$ must have the same symmetry as the background structure. The class of all physical solutions to some set of field equations thus falls into equivalence classes under the symmetry group of the background metric. But, in the background-dependent case, the members of each equivalence class can be physically distinguished from each other by  those further specifications that enable a distinction to be made between the points on the orbits of the symmetry group.

In the case of space-times with a fixed background structure, the background metric will have a symmetry group that may be a finite (up to ten) parameter Lie group or the trivial group. We refer to the latter class of metrics as {\em generic} space-times. 
The points of generic space-times are completely individuated (at least locally) by the values of four invariants of the Riemann tensor. \footnote{Even if the Ricci tensor vanishes, there will  be four such non-vanishing invariants in the generic case.} If there is a (non-zero) finite-parameter Lie group, some further specification of the points on the orbit(s) of the group is needed in order to individuate them. 

In the important case of special-relativistic field theories, the specification of a fibration consisting of parallel timelike geodesic (''straight'') world-lines and the corresponding orthogonal foliation of spacelike (''flat'') hyperplanes singles out  a particular inertial frame of reference in Minkowski space-time. Choice of a spatial origin world line of the flibration and a temporal origin hyperplane of the foliation, together with units of space and time, will then completely individuate each point of Minkowksi space. 
The individuation of the points of Minkowski space by means of choice of an inertial frame of reference, together with a specification of spatial and temporal origins and units, enables us to distinguish physically between members of each equivalence class of solutions under the Poincar{\'e} group. The Coulomb field of a charge $q$ centered at the origin of one inertial frame belongs to the same equivalence class as the Coulomb field of a charge $q$ centered at the origin of another inertial frame moving with velocity $v$ with respect to the first; but with respect to the first frame the two are quite different physical fields.

\subsection{Blocking the Formulation of the Hole Argument}

For theories that are generally covariant, in order to avoid the formulation of this generalized version of the hole argument we define a quotient space of classes of models, so that each class of models corresponds to a (unique) physical model.

Einstein's idea implies that if you drag-along the metric field, you drag all the physical properties of space-time points along with them. Therefore, a whole class of 
diffeomorphically related metrics corresponds to a unique physical gravitational field.

Based on the construction of the quotient space of classes of models of a theory, we obtain a 1:1 correspondence between a point in the quotient space ( i.e., a class of diffeomorphically related mathematical models) and a (unique) physical model of the theory.

Let us consider a background independent theory $\mathcal T$ defined on a manifold $M$ i.e. a rule for selecting a set of cross-sections of a natural bundle $(E\stackrel{\pi}\to M)$.

Denote $\Gamma(E;M)$ the collection of all cross-sections of $\pi$. The group of base diffeomorphisms $Diff(M)$ acts as a transformation group on $\Gamma(E;M)$ by
carrying-along cross-sections of $\pi$: for all $\phi\in Diff(M)$ and $\sigma\in \Gamma(E;M)$ the action map is defined by $(\phi, \sigma)\longmapsto \phi^{*}(\sigma) $. 

For a fixed model $\sigma$, $\mathcal{O}_{g}=\{\phi^{*}(g)|\quad \phi\in Diff(M)\}$ is the orbit through $\sigma$. Two cross-sections $\sigma_1$ and $\sigma_2$ are on the same orbit, if and only if $\phi^{*}(\sigma_1)=\sigma_2$ for some  $\phi\in Diff(M)$. The action of $Diff(M)$ on $\Gamma(E;M)$ partition $\Gamma(E;M)$ into (disjoint) diffeomorphically equivalent classes of cross-sections.

The space $\mathcal{Q}(M)$ of all diffeomorphically-equivalent classes of cross-sections on $M$ 
is $\Gamma(E;M)/Diff(M)$. A physical model corresponds to a point in this quotient space.  

The space $\mathcal{Q}(M)$ usually is singular and not a manifold and the infinite-dimensional analysis is quite complicated. In fact, only if the action of the Lie group is free i.e. all isotropy subgroups $\mathcal{H}_{\sigma}=\{\phi \in Diff(M) | \phi^{*}(\sigma)=\sigma\}$ are trivial, the resulting orbit space bears a manifold structure and $(\Pi:\Gamma(E;M)\longrightarrow \mathcal{Q}(M))$ forms a principal fiber bundle. 
More often, the orbit space $\mathcal{Q}(M)$ admits a stratification into manifolds. The existence of such a stratification is usually shown by proving the existence of slices at every point for the group action. \footnote{For details about the slice theorem, see \cite{Palais}}

The projection map $\Pi:\Gamma(E;M)\longrightarrow \mathcal{Q}(M)$ identifies all diffeomorphically-equivalent models to a single diffeomorphically-equivalence class. The image of $\Pi(\sigma)=[\sigma]$ represents the physical model defined by $\sigma$.

\section{Conclusion}

In our approach we start with a fibered manifold representing the metric and
perhaps other physical fields, rather than the base manifold. This way it becomes 
difficult to forget that, in general relativity, space-time is no more than ``a structural quality of the field.'' This agrees with one of Einstein's intuitions about general relativity:

\begin{quote}
On the basis of the general theory of relativity, space as opposed to 
{\em what fills space} ... has no separate existence … 
If we imagine the gravitational field, i.e., the functions $g_{ik}$ 
to be removed, there does not remain a space of the type (of special relativity), 
but absolutely nothing, and also not {\em topological space}. ... 
There is no such thing as an empty space, i.e., a space without field. 
Space-time does not claim existence on its own, but only as a structural quality of the 
field. ({\em Albert Einstein }) \cite{Stachel1986}, pp. 1860, \cite{Einstein1952}, pp. 155.
 
\end{quote}

In a backgound independent theory ( e.g. general-relativistic theories), if we start from some natural bundle, the base points may be characterized as such independently of the particular relations in which they stand; but they are entirely individuated in terms of the relational structure given by cross-sections of some fibered manifold. 

In our invariant formulation, the hole argument can be easily further generalized to the case of sets and relation with some important implications to problems in quantum gravity. \cite{IftimeStachel}. Elementary particles are similarly individuated by their position in a relational structure and this suggests the following viewpoint: Since the basic building blocks of any model of the universe, the elementary particles and the points of space-time, are individuated entirely in terms of the relational structures in which they are embedded, only ''higher-level'' entities constructed from these building blocks can be individuated independently. Therefore, the following principle of generalized covariance should be a requirement on any fundamental theory: The theory should be invariant under all permutations of the basic elements, out of which the theory is constructed.

\section{Acknowledgements} Mihaela Iftime would like to thank Gennadi Sardanashvily, Vincent Moncrief and Norma Chase for suggestions. John Stachel thanks Abhay Ashtekar and Oliver Pooley for their comments of an earlier version of this paper.

\end{document}